\documentclass[preprintnumbers,showpacs, showkeys, twocolumn,secnumarabic,amssymb,amsmath,amsfonts,aps, pra]{revtex4}

\usepackage{graphicx}

\usepackage{amsmath}
\begin{document}

\newcommand{\non}{\nonumber}
\newcommand{\be}{\begin{equation}}
\newcommand{\ee}{\end{equation}}
\newcommand{\bq}{\begin{eqnarray}}
\newcommand{\eq}{\end{eqnarray}}
\newcommand{\lps}{\langle}
\newcommand{\rps}{\rangle}

\title{Microcanonical Entropy and Dynamical Measure of Temperature for Systems with 
Two First Integrals}

\date{\today}

\author{Roberto Franzosi}
\affiliation{C.N.I.S.M. UdR di Firenze, Dipartimento di Fisica,
Universit\`a degli Studi di Firenze, Via Sansone 1,
I-50019 Sesto Fiorentino and
I.P.S.I.A. C. Cennini, Via dei Mille 12/a, I-53034 Colle di Val d'Elsa (SI),
Italy.}

\begin{abstract}
We consider a generic classical many particle system described by an autonomous
Hamiltonian $H(x^{_1},\ldots,x^{_{N+2}})$ which, in addition, has a conserved
quantity $V(x^{_1},\ldots,x^{_{N+2}})=v$, so that the Poisson bracket $\{H,V \}$
vanishes. We derive in detail the microcanonical expressions for entropy and
temperature. We show that both of these quantities depend on multidimensional
integrals over sub-manifolds given by the intersection of the constant energy
hyper-surfaces with those defined by $V(x^{_1},\ldots,x^{_{N+2}})=v$.
We show that temperature and
higher order derivatives of entropy are microcanonical observable that, under
the hypothesis of ergodicity, can be calculated as time averages of suitable
functions. We derive the explicit expression of the function that gives the
temperature.

\end{abstract}
\maketitle

For an isolated many-body classical system, the ergodicity makes equivalent
thermodynamics and dynamics. Thus, in these cases, one can measure thermodynamic
quantities, like temperature and specific heat, as temporal averages of suitable
functions along almost each trajectory of a given system, or, equivalently, as
an integral over its phase space.
The opportunity to pass from the dynamics to the microcanonical-thermodynamics
and vice versa, gives the possibility to choose the smarter way
to measure a given quantity.
Very often the calculation of thermodynamic quantities in the microcanonical
ensemble is an impracticable issue, thus one is forced to recur to the canonical
ensemble, where these measures are more easily performed by resorting to numeric
simulations, e.g. by Monte Carlo method.
Of course, for a given system this step is performable only if there is equivalence
between canonical and microcanonical ensembles in the thermodynamic limit.
Furthermore, it is worth mentioning that in many cases, the convergence of
thermodynamic quantities to their asymptotic values, is much faster if the averages
are computed along the dynamics, rather than by an important sampling of the canonic
phase-space. For these cases therefore, the dynamics is preferred respect to
statistics.
Furthermore, only for systems described by stable and temperated
inter-particle interaction potentials is guaranteed the equivalence of the statistical
ensembles in the thermodynamic limit.
Nowadays, several of the most intriguing
challenge for modern science, deals with systems of size intermediate between the
macroscopic and the microscopic scales. Systems like polymers, DNA-helix,
proteins, nanosystems, are large enough to allow a statistical treatment, but are
absolutely far from the thermodynamic limit. Thus, for these systems, ensemble equivalence
is hardly verified and one has no option but performing dynamical averages.

It is in this same spirit that Rugh, in \cite{rugh}, has presented a dynamical approach
for measuring the temperature of a Hamiltonian system in the microcanonical ensemble.
He has shown that  for an ergodic classical system, which has only one conserved quantity,
i.e. the energy, the inverse temperature $1/T$, can be calculated as a temporal average of
a suitable functional along the dynamics.
The Rugh's seminal work has stimulated several papers
\cite{rugh1,otter,rickayzen}
which have aroused widespread interest, especially among those
who simulate the properties of liquids \cite{liquids}.
The calculation given by Rugh in Ref. \cite{rugh} provides a microcanonical
definition of temperature that allows its measure also in systems with nonstandard
Hamiltonians. Nevertheless, this calculation works for systems with only one
conserved quantity, i.e. the total energy. \textit{In the present paper we extend the
calculation of entropy and of microcanonical temperature  to the case of Hamiltonian
systems with two first integrals of motion.}

In the present paper we consider a classical system of $(N+2)$ degrees of freedom
(with $N>0$), which is described by an autonomous Hamiltonian $H$, and which has a
further independent conserved quantity $V$, such that $\{H,V \} = 0$.
\textit{We derive in detail the microcanonical expressions for entropy,
i.e. the expression of the microcanonical invariant measure, and the temperature,
moreover we give a formula to derive, recursively, all order of derivatives of
the entropy}.
We show that entropy and temperature depend on multidimensional
integrals over sub-manifolds given by the intersection of the constant energy
hyper-surfaces with those defined by $V(x^{_1},\ldots,x^{_{N+2}})=v$.
In particular, we show that temperature and higher order derivatives of entropy
are microcanonical observable that, under the hypothesis of ergodicity, can be
calculated as time averages of suitable functions.
In Ref. \cite{rugh2} it has been studied the microcanonical ensemble
of a classical system, whose Hamiltonian is parameter
dependent and in presence of other first integrals and it has been showed a
method, alternative to the present one, that allows one to obtain the first
derivative of entropy respect to the conserved quantities.
The aim of the present paper is to derive \textit{explicitly}  the functional
by means of which the temperature can be calculated as a microcanonical average,
in the case of a generic classical system, described by a many-body Hamiltonian
with one further conserved quantity. Furthermore, our method allowed us to derive an
iterative formula that gives the derivatives of $S(E)$ of all orders, for
this class of systems. By this formula, one can measure more general quantities
like, e.g. the specific heat.

Let ${H}(\mathbf{x},x^{_{N+1}},x^{_{N+2}})$ be a classical Hamiltonian
describing an autonomous many-body system whose coordinates and
canonical momenta $(q_1,p_1,\dots)$ are represented as $(N+2)$-component vectors
$(\mathbf{x},x^{_{N+1}}, x^{_{N+2}}) \in \mathbb{R}^{_{N+2}}$, and let 
$V(\mathbf{x},x^{_{N+1}},x^{_{N+2}})$ be a further conserved quantity which
is in involution with $H$. We shall assume that $V$ is a smooth function of
the coordinates. The system's motion takes place on the manifolds
$\mathcal{M} =\Sigma_E \cap V_u$, where the $\Sigma_E=
\{(\mathbf{x},x^{_{N+1}},x^{_{N+2}}) \in\mathbb{R}^{_{N+2}}|
{H}(\mathbf{x},x^{_{N+1}},x^{_{N+2}}) =E\}$
are energy level sets and $V_u=
\{(\mathbf{x},x^{_{N+1}},x^{_{N+2}}) \in\mathbb{R}^{_{N+2}}|
V(\mathbf{x},x^{_{N+1}},x^{_{N+2}}) =u\}$ are subsets of $\mathbb{R}^{_{N+2}}$
where $V$ is constant.
Among the equivalent expressions allowed for the microcanonical entropy $S(E)$, the surface
entropy~\cite{khinchin}
\begin{multline}
 S(E) = \ln\! \int \! d^{N}\mathbf{x} d x^{_{N+1}} d x^{_{N+2}}
\delta({H}(\mathbf{x},x^{_{N+1}},x^{_{N+2}}) -E) \times \\
\delta({V}(\mathbf{x},x^{_{N+1}},x^{_{N+2}}) -u) 
\label{entropy2}
\end{multline}
has an interesting and useful geometric interpretation
that we shall derive, following the calculation shown in the chapter on the
theory of surfaces of Ref. \cite{DNF}. Consistently with Ref. \cite{DNF},
we shall assume the level sets of ${H}$ and $V$ to be non-singular hyper-surfaces.
Even if the energy level surfaces (or the level sets of $V$) in general constitute
a singular foliation, thus for some values of $E$ (or $u$) the energy surface
(or $V_u$) is not a differential manifold, for generic values of $E$ (or $u$) this is
not an issue.
{For a generic point $x_0\in \mathcal{M}$ of a non-singular level-set of ${H}$ and $V$,
$\nabla{H} (x_0)\neq 0$ and $\nabla{V} (x_0)\neq 0$. Furthermore, since $H$ and
$V$ are in involution, almost everywhere $\nabla{H} (x_0)$ and $\nabla{V} (x_0)$ are
independent vectors. Thus, in a neighborhood of $x_0$ we can suppose
of reorder the coordinate indices in such a way that
${\partial H}/{\partial x^{_{N+1}}} {\partial V}/{\partial x^{_{N+2}}}
-{\partial V}/{\partial x^{_{N+1}}} {\partial H}/{\partial x^{_{N+2}}} \neq 0$
for each $x$ of the neighborhood.
Now, each non-singular manifold $\mathcal{M}$, can be partitioned by a
family $\mathcal{F}$ of not overlapping subsets \footnote{More precisely
we can make use of the partition-of-unity of the level set.}. With the further
condition that, in each subset we can reorder the coordinate indices, as said
above, so that ${\partial H}/{\partial x^{_{N+1}}} {\partial V}/{\partial x^{_{N+2}}}
-{\partial V}/{\partial x^{_{N+1}}} {\partial H}/{\partial x^{_{N+2}}} \neq 0$
 everywhere
in the subset.}
{ 
Now, by using the first $N$ coordinates we give a parametric
description of the same subset.
Thus, in each of the subsets of $\mathcal{F}$ we can choose}
$f^\alpha = id$ for $\alpha = 1,\dots,N$, and let us set
$g(\textbf{y}) := f^{_{N+1}}(\textbf{y})$ and  $h(\textbf{y}) := f^{_{N+2}}(\textbf{y})$,
where $\textbf{y} \in \mathbb{R}^{_N}$.
The metric induced by $\mathbb{R}^{_{N+2}}$ on $\mathcal{M}$ results
\begin{equation}
\eta_{\mu \nu} = \delta_{\mu \nu} +
{\partial_\mu g}{}
{\partial_\nu g}{}+
{\partial_\mu h}{}
{\partial_\nu h}{} \, ,
\label{etamunu1}
\end{equation}
where $\partial_\alpha \bullet =\partial \bullet/ \partial x^\alpha$, whereas the its
determinant can be derived by straightforward calculations and it results
\begin{multline}
\eta = 1 + \sum^N_{\alpha=1} \left[  \left( 
{\partial_\alpha g}{} \right)^2 + \left( 
{\partial_\alpha h}{} \right)^2 \right] + \\
\sum^N_{{\mu,\nu=1}\atop{\mu < \nu}} \left( 
{\partial_\mu g}{}
{\partial_\nu h}{} -
{\partial_\mu h}{}
{\partial_\nu g}{}
 \right)^2 \, .
\label{eta2}
\end{multline}
The derivatives ${\partial_\alpha g}{}$, ${\partial_\alpha h}{}$, can be expressed
as follows
\begin{eqnarray}
\partial_\alpha g&=
\left[ \partial_{^{N+2}} V
       \partial_\alpha H -
       \partial_{^{N+2}} H
       \partial_\alpha V \right]/D \label{dg}\\
\partial_\alpha h &=
\left[ \partial_{^{N+1}} H
       \partial_\alpha V -
       \partial_{^{N+1}} V
       \partial_\alpha H \right]/D \label{dh}\, ,
\end{eqnarray}
where $D =\! {\partial_{_{N+1}} H}{} {\partial_{_{N+2}} V}{}
-{\partial_{_{N+1}} V}{} {\partial_{_{N+2}} H}{}$
\footnote{{ Each subset of $\mathcal{F}$ can be given in a parametric form of $N$
coordinates if $D(x) \neq 0$ for each $x$ in the subset.}}.
From the expression above we derive the sub-manifold volume form
\begin{equation}
 d \tau = d^N \textbf{x} \sqrt{\eta} = d^N \textbf{x}
\dfrac{W}{D} \, ,
\end{equation}
where
\begin{equation}
W = 
\left[
\sum^{N+2}_{{\mu , \nu=1}\atop{\mu < \nu}}
\left( 
\dfrac{\partial H}{\partial x^{\mu}} \dfrac{\partial V}{\partial x^{\nu}} 
-
\dfrac{\partial H}{\partial x^{\nu}} \dfrac{\partial V}{\partial x^{\mu}} 
\right)^2 
\right]^{1/2}
\, .
\label{eta2new}
\end{equation}
On the other hand, expression (\ref{entropy2}) can be cast in the following form.
For each point $\textbf{x}$ we can introduce the following variables change
\begin{equation}
 y^1 
= h^1_{\textbf{x}} (x^{_{N+1}}, x^{_{N+2}}) 
\, ,\label{y1} \
 y^2 
= h^2_{\textbf{x}} (x^{_{N+1}}, x^{_{N+2}}) \, ,
\end{equation}
with the inverse transformations
\begin{equation}
 x^{_{N+1}} 
= G^1_{\textbf{x}} (y^1, y^2) 
\, , \
 x^{_{N+2}} 
= G^2_{\textbf{x}} (y^1, y^2) \, ,
\label{x2}
\end{equation}
where $h^1_{\textbf{x}} (x^{_{N+1}}, x^{_{N+2}}) = {H}(\mathbf{x},x^{_{N+1}},x^{_{N+2}}) -E$,
$h^2_{\textbf{x}} (x^{_{N+1}}, x^{_{N+2}}) = {V}(\mathbf{x},x^{_{N+1}},x^{_{N+2}}) -u$,
and $G^1_{\textbf{x}} (y, z)$ and $G^2_{\textbf{x}} (y, z)$ are the respective inverse
transformations. From now on we will suppress the sub-index $_{\textbf{x}}$.
After these definitions Eq. (\ref{entropy2}) can be expressed as follows
\[
 S(E) = \ln \int d^{N}\mathbf{x} d y^1 d y^2
\delta(y^1) \delta(y^2) \vert J(y^1,y^2) \vert \, ,
\]
where $\vert J(y^1,y^2) \vert$ is the determinant of the Jacobian matrix
$\partial (x^{_{N+1}},x^{_{N+2}})/\partial (y^1,y^2)$.
The inverse of the Jacobian matrix can be derived by Eqs. (\ref{y1})-(\ref{x2}) and
it results
\begin{eqnarray}
 \partial_1 G^1 &=& - {\partial_2 h^2}/{D} = - {\partial_{^{N+2}} V}/{D} \, ,\\
 \partial_1 G^2 &=&   {\partial_1 h^2}/{D} =   {\partial_{^{N+1}} V}/{D} \, ,\\
 \partial_2 G^1 &=&   {\partial_2 h^1}/{D} =   {\partial_{^{N+2}} H}/{D} \, ,\\
 \partial_2 G^2 &=& - {\partial_1 h^1}/{D} = - {\partial_{^{N+2}} H}/{D} \, .
\end{eqnarray}
Thus, the microcanonical entropy  \footnote{\label{noteSrugh}
The analogous expression valid for systems with only one first
integral, derived in \cite{khinchin}, is the following
$
S(E) = \ln \int_{\Sigma_E}  {d \sigma}/{\Vert \bigtriangledown H \Vert} \, .
$}
results
\begin{equation}
 S(E) = \ln \int_{\mathcal{M}} d^{N}\mathbf{x}
\dfrac{1}{D}
= \ln \int_{\mathcal{M}}
\dfrac{d \tau}{W} \, .
\label{entropy-geometry-new}
\end{equation}
{ \textit{ It is worth emphasizing that the invariant measure is independent
from the partition of $\mathcal{M}$, since $W$ is invariant under exchange
of the indices of the coordinates}.} 

In order to derive the temperature in the microcanonical ensemble, according
to the definition $T(E)=(\partial S (E)/\partial E)^{-1}$, we shall use the
following generalization \footnote{The Federer-Laurence derivation formula 
\cite{federer,laurence} is
$
\partial^k (\int_{\Sigma_E}  \psi
 d \Sigma )/\partial E^k  = \int_{\Sigma_E} A^k\left( \psi\right)
 d \Sigma
$,
where $A(\bullet) = 1~/~\|~\bigtriangledown~H~\|~\bigtriangledown
 \left( \bigtriangledown H/\|\bigtriangledown H \| \bullet \right)$.
} of the Federer-Laurence derivation formula \cite{federer,laurence,TH1,TH2,PettiniBook}.
The flux $\Phi_\xi$ with a non-vanishing component in the direction
orthogonal to the constant energy hyper-surfaces, but tangent to
the level hyper-surfaces of $V$, can be defined by the vector
$
 \xi = n^{_H} - ( n^{_H} \cdot n^{_V}  ) n^{_V} 
$,
where $n^{_H}={\bigtriangledown H}/{\Vert \bigtriangledown H \Vert }$ and
$n^{_V}={\bigtriangledown V}/{\Vert \bigtriangledown V \Vert }$.
Let us define $n^{_\xi} = \xi/\Vert \xi \Vert$, so
the generalized derivation formula results
\begin{equation}
\dfrac{\partial^k }{\partial E^k} \int_\mathcal{M} d \tau \psi
  = \int_\mathcal{M} d \tau A^k\left( \psi \right)
  \, ,
\label{federer-laurence}
\end{equation}
where \footnote{Here, and in the following 
\[
n^{_V} \cdot (n^{_V} \cdot 
 \bigtriangledown)(n^{_\xi}) 
= \sum_{j,k} n^{_V}_j  n^{_V}_k \partial_k(n^{_\xi}_j) \, .
\]
}
\begin{equation}
 A (\bullet) = \frac{1}{\bigtriangledown H \cdot n^{_\xi}} \left[ \bigtriangledown 
\left( n^{_\xi} \bullet \right) -
\bullet n^{_V} \cdot (n^{_V} \cdot  \bigtriangledown)(n^{_\xi})\right] \, .
\label{Anew}
\end{equation} 
The proof of the extention of the Federer-Laurence theorem to varieties of codimension two,
is rather complicated and lengthy. Furthermore it is outside of the main motivation of
the present paper, thus it will be given in a further paper \cite{kfirs}.
After Eqs. (\ref{niceformula1}) and (\ref{niceformula2}), we obtain that the inverse temperature
is given by
\begin{equation}
 \dfrac{1}{T(E,V)} = 
\left\langle  
  \Phi(x)
\right \rangle_{\mu c} \, ,
\label{niceformula1}
\end{equation}
where 
\begin{multline}
\Phi(x) \! = \! 
		\dfrac{W}{ \bigtriangledown H \cdot n^{_\xi} } 
		\left[ 
		 \bigtriangledown \left( 
\frac{n^{_\xi}}{W} \right)
- \frac{(n^{_V} \cdot \bigtriangledown) \left( n^{_\xi} \right) }{ W} 
\cdot
n^{_V}
		\right] \, ,
\label{niceformula2}
\end{multline}
and $\left\langle \right \rangle_{\mu c}$ stands for the microcanonical average.

When $(\mathbb{R}^{_{N+2}},H)$ is ergodic with respect to the Liouville measure
 $d \tau/W$ restricted to a nonsingular manifold $\mathcal{M}$ for almost every
initial condition $x (0) \in \mathcal{M}$ one has
\begin{equation}
 \dfrac{1}{T(E,V)} = \lim_{s \to \infty} \frac{1}{s} \int^s_0 d s^\prime
\left[ 
\Phi (x(s^\prime))
\right] \, .
\end{equation}

\vskip 0.25cm
{\em Simple geometric applications.} \\ \noindent
\textit{i)}\quad As a simple application we shall derive the invariant metric $\eta$ in the
case of a simple form for $H$ as that one of a four-dimensional
hypersphere $H = x^2 + y^2 + z^2 + w^2$ of unit radius, and with a condition
given by the hyper-plain $V=z+w=0$. By direct calculations we derive
$\partial_x H = 2 x$, $\partial_y H = 2 y$, $\partial_z H = 2 z$, $\partial_w H = 2 w$,
thus it results $D = \partial_z H \partial_w V - \partial_z V \partial_w H = 2 (z-w)$.
With the notations introduced above, Eqs. (\ref{dg}) and (\ref{dh}) become
$\partial_x g = x/(z-w)$, $\partial_y g = y/(z-w)$, 
$\partial_x h = - x/(z-w)$, and $\partial_y h = - y/(z-w)$, thus the invariant metric
results
\begin{equation}
 \eta = 1 + 2 \left[  \left( \frac{x}{z-w}\right)^2 
+ \left( \frac{y}{z-w}\right)^2 \right] \, ,
\label{check}
\end{equation}
and is defined for $D\neq0$ ($z \neq w$) that is on the subsets $$\{ (x,y, 
\pm \sqrt{\frac{1-x^2-y^2}{2}}, \mp \sqrt{\frac{1-x^2-y^2}{2}})| x^2~+~y^2~<~1\}.$$
The sum of the integrals of $\sqrt{\eta}$ upon
these subsets brings to the result $4\pi$ which is the right hyper-surface volume.
This solution can be checked noting that in the original problem the condition
$V=z+w=0$ can be lifted by a suitable change of variables.
Let us introduce the variables $s = (z-w)/\sqrt{2}$ and $t = (z+w)/\sqrt{2}$. $H$ and
$V$ can be expressed in the new variables as $H = x^2 + y^2 + s^2 + t^2 =1$ and $V=t=0$.
Thus the metric $\eta$ is that one of the three-dimensional unitary sphere
$x^2 + y^2 + s^2 =1$ that is (see \cite{DNF}) $\eta = 1 + (x/s)^2 + (y/s)^2$.
This indeed is the expression
(\ref{check}). The sum of the integrals of $\sqrt{\eta}$ upon the two hemispheres,
$s> 0$ and $s< 0$, gives the right value $4 \pi$.
\vskip 0.25cm
\noindent
\textit{ii)}\quad Let us now consider a simple case where all the geometric quantities
can be explicitly calculated in order to check the Eqs. (\ref{federer-laurence})
and (\ref{Anew}).
We shall consider $H=x_1^2+x_2^2+x_3^2=E$, a sphere of radius $\sqrt{E}$ in three dimensions
and $V=x_3/\sqrt{x_1^2+x_2^2}=u$, that is a cone with an angle $\arctan(u)$ at the vertex.
In this case $\mathcal{M}$ is a circle of radius $a=\sqrt{E/(1+u^2)}$.
\vskip 0.25cm
\noindent
If we choose $\psi=1$,
we get easly 
\[
 \dfrac{\partial}{\partial E }
 \int_\mathcal{M} d \tau = \dfrac{\partial}{\partial E } 2 \pi a = \dfrac{\pi a}{E}
  \, .
\]
The terms that appear in Eq. (\ref{Anew}) result $\bigtriangledown H \cdot n^{_\xi} = 2 \sqrt{E}$,
$\bigtriangledown n^{_\xi} = 2/\sqrt{E}$ and
$n^{_V} \cdot (n^{_V} \cdot  \bigtriangledown)(n^{_\xi}) = 1/\sqrt{E}$. Thus $A(1)=1/(2 E)$ and
consequently $\int_\mathcal{M} d \tau A\left( 1 \right)=  \pi a / E$.
\vskip 0.25cm
\noindent
By choosing $\psi=1/W$ we get $W=2E/a^2$, and 
\begin{equation}
 \dfrac{\partial}{\partial E }
 \int_\mathcal{M} \dfrac{d \tau}{W} = \dfrac{\partial}{\partial E } 
\dfrac{a^2}{2 E} 2 \pi a = \dfrac{\pi}{2 (1+u^2)^{{3/2}} \sqrt{E}}
  \, .
\label{second} 
\end{equation}
Again, Eq. (\ref{Anew}) contains $\bigtriangledown H \cdot n^{_\xi} = 2 \sqrt{E}$,
$\bigtriangledown (n^{_\xi}/W) = a^2/{E^{{3/2}}}$ and
$n^{_V} \cdot (n^{_V} \cdot  \bigtriangledown)(n^{_\xi})/W = a^2/(2 E^{{3/2}})$.
Thus $A(1/W)=a^2/(4 E^2)$ and
consequently $\int_\mathcal{M} d \tau A\left( 1/W \right)=  \pi a^3 / (2E^2)$
which indeed coincides with result (\ref{second}).

\vskip 0.25cm
{\em Dynamical system.} \\ \noindent
Let us now consider a lattice system described by the Hamiltonian
\begin{equation}
 { H} = \frac{\nu}{8} \sum_{m} (p^2_m + q^2_m)^2 -
 \sum_{m} (p_{m} p_{m+1} +q_{m} q_{m+1}) \,
\label{HBH}
\end{equation}
and the usual Poisson brackets $\{ q_m,p_n \} = \delta_{m,n}$ for $n,m=1,\ldots,M$, with
periodic boundary conditions. The dynamics generated by this Hamiltonian conserves
the quantity $N = \sum_m (q^2_m + p^2_m)/2$, thus the Eqs. (\ref{niceformula1}) and
(\ref{niceformula2}) are in order to calculate the microcanonical temperature.
The ground-state is achieved by solving the equation $\delta ({ H} - \chi N) = 0$
in which the Lagrangian multiplier $\chi$ has been introduced to take in account
the conservation of $N$.
By direct calculations, we got the solution ${q_0}_m = \sqrt{2 N/M}:=q_0$,
${p_0}_m = 0$ and $\chi = \nu N/M -2$
\footnote{Another and equivalent solution is given by
${q_0}_m = 0$, ${p_0}_m = \sqrt{2 N/M}$ and $\chi = \nu N/M -2$.}.
Small fluctuations around this ground-state correspond to $T \gtrsim 0$, in the
following we show that this prevision is verified by Eq. (\ref{niceformula1}) and
(\ref{niceformula2}),
whereas the formula for $1/T$ given in \cite{rugh}, that holds in the case of
systems with only one first integral (energy), does not work.
The reason of this comparison is to show that the equation derived in \cite{rugh},
which is valid in the case of system with only one first integral, cannot
be used in the case of systems with more than one first integrals
{{ In Ref. \cite{rugh2} has been derived an equation similar to
Eqs. (\ref{niceformula1}) and (\ref{niceformula2}). The inverse temperature $1/T$ can be derived by
Eqs. (12) and (17) of Ref. \cite{rugh2} after having found two vectors, $X_0$
and $X_1$, such that $d{H}(X_0) = 1$, $d{H}(X_1) = 0$, $d {V}(X_0) = 0$
and $d {V}(X_1) = 1$. One can use $X_0=c_{VV} \nabla H /d - c_{VH} \nabla V/d$ and
$X_1=-c_{VH} \nabla H /d + c_{HH} \nabla V/d$, where $c_{HH} = \| \nabla H \|^2 $,
$c_{VV} = \|\nabla V \|^2$, $c_{VH} = \nabla V \nabla H $, and $d=c_{HH} c_{VV}
- c^2_{VH}$. Thus it results $1/T = \left\langle \nabla \cdot X_0 \right\rangle_{\mu c} $.
The latter expression seems inequivalent to Eqs. (\ref{niceformula1}) (\ref{niceformula2}).
Indeed, it does not contain $1/W$ which is related to the invariant measure.
It is worth emphasizing that an analogous term $1/\| {H} \|$, which is related to the
microcanonical measure of a system with one first integral, indeed appears in the formula
for $1/T$ derived in \cite{rugh}.
In any case a comparison between these two equations, would require a
numerical simulation, but this is out of the aim of the present paper.}}
By expanding ${ H}$ in terms of the displacements $Q_n$ and
$P_n$ from the minimum points $q_n = q_0$ and $p_n = 0$, and by calculating the
terms appearing in (\ref{niceformula1}) and (\ref{niceformula2})
in the limit $Q_n,P_n \to 0$, after some boring algebra we got
the correct low-energy temperature, being $1/T = \left\langle \Phi \right\rangle_{\mu c}
\to \infty$.

It is worth emphasizing that by using the formula derived in Ref. \cite{rugh}
$
{1}/{T} = \left\langle
\nabla \left( { \nabla H}/{\Vert \nabla H \Vert^2} \right) 
\right\rangle_{\mu c}
$,
which is correct for systems with only one first integral,
one obtains the erroneous result: $T =
\left[ \left( \frac{\nu}{2}  q^2_0 -2\right)q_0\right]^2/(2 \nu q^2_0)$.

In conclusion, we have presented a dynamical approach for measuring the temperature
of a Hamiltonian system with two first integrals in the microcanonical ensemble.
The formula we have derived allows
one to measure the inverse temperature as a time-average, instead of as an average
over the phase-space, also in the case of systems with two first integrals.
{ {\em Furthermore, by Eq. (\ref{federer-laurence}) higher orders of derivatives of
$S(E)$ can be obtained. Therefore the method here presented allows one to measure
g.e. the specific-heat.}}

I'm indebted with Prof. M. Pettini, Prof. G. Vezzosi and Prof. R. Livi for the
usefull discussions.
\hfill

\end{document}